\documentclass[10pt,journal,compsoc]{IEEEtran}

\ifCLASSOPTIONcompsoc
  \usepackage[nocompress]{cite}
\else
  \usepackage{cite}
\fi
\usepackage{cite}
\usepackage[numbers,sort&compress]{natbib}
\usepackage{graphicx}
\usepackage{amsmath}
\usepackage{multirow}
\usepackage{float}
\usepackage{bm}
\usepackage{amsfonts}
\usepackage{cite}
\usepackage{hyperref}

\ifCLASSINFOpdf
\else
\fi
\begin{document}
%
\title{Emotion Selectable End-to-End  Text-based Speech Editing}
%
%
%
%

\author{Tao~Wang,
Jiangyan~Yi, ~\IEEEmembership{Member,~IEEE,}
        Ruibo~Fu,~\IEEEmembership{Member,~IEEE,}
         Jianhua~Tao,~\IEEEmembership{Senior~Member,~IEEE,}
   ~Zhengqi~Wen, and Chu Yuan Zhang 
\IEEEcompsocitemizethanks{\IEEEcompsocthanksitem T. Wang is with the National Laboratory of Pattern Recognition, Institute of
Automation, Chinese Academy of Science, Beijing 100190, China, and also
with the School of Artificial Intelligence, University of Chinese
Academy of Sciences, Beijing 100190, China. \protect\\
E-mail: tao.wang@nlpr.ia.ac.cn
\IEEEcompsocthanksitem J. Yi, R. Fu, J. Tao, Z. Wen and C. Zhang are  with the National Laboratory of Pattern Recognition,
Institute of Automation, Chinese Academy of Sciences, Beijing 100190, China
(e-mail: \{jiangyan.yi, ruibo.fu,  jhtao, zqwen, chuyuan.zhang\}@nlpr.ia.ac.cn). Corresponding Author: Jiangyan Yi, Ruibo Fu,    Jianhua Tao.}
\thanks{Manuscript received April 19, 2005; revised August 26, 2015.}}

%
%

\markboth{Journal of \LaTeX\ Class Files,~Vol.~14, No.~8, August~2015}%
{Shell \MakeLowercase{\textit{et al.}}: Bare Advanced Demo of IEEEtran.cls for IEEE Computer Society Journals}
%



\IEEEtitleabstractindextext{%
\begin{abstract}
Text-based speech editing allows users to edit speech by intuitively cutting, copying, and pasting text to speed up the process of editing speech. In the previous work, CampNet (context-aware mask prediction network) is proposed to realize text-based speech editing, which significantly improves the quality of editing speech. This paper aims at a new task: adding emotional effect to the editing speech during the text-based speech editing to make the generated speech more expressive. To achieve this task, we propose Emo-CampNet (emotion CampNet), which can provide the option of emotional attributes for the generated speech in text-based speech editing and has the one-shot ability to edit any speaker's speech.  Firstly, we propose an end-to-end emotion selectable  text-based speech editing model. The key idea of the model is to control the emotion of generated speech by introducing additional emotion attributes based on the context-aware mask prediction network. Secondly, to prevent the emotion of the generated speech from being interfered by the emotional components in the original speech, a neutral content generator is proposed to remove the emotion from the original speech, which is optimized by the generative adversarial framework. Thirdly, two data augmentation methods are proposed to enrich the emotional and pronunciation information in the training set, which can enable the model to edit the speech of any speaker.

The experimental results \protect\footnotemark{show} that: 1) Emo-CampNet can effectively control the emotion of the generated speech in the process of text-based speech editing; And can edit the speech of any speaker. 2) Detailed ablation experiments further prove the effectiveness of emotional selectivity and data augmentation methods. 
\end{abstract}

}

\maketitle

\IEEEdisplaynontitleabstractindextext

%
\IEEEpeerreviewmaketitle

\ifCLASSOPTIONcompsoc
\IEEEraisesectionheading{\section{Introduction}\label{sec:introduction}}
\else
\section{Introduction}
\label{sec:introduction}
\fi

%
%
%
%
\IEEEPARstart{W}{ith}  the rapid development of the internet, there are all kinds of media for us to learn, entertain and communicate. These media are closely related to the generation of audio. Text-based speech editing \cite{jin2018speech,morrison2021context,tan2021editspeech,wang2022campnet,bai20223}, that is, modifying the audio by directly editing the transcript, will bring great convenience to the audio generation process. For example, content creators can quickly edit transcripts using familiar word processing operations, such as cut, copy, and paste, and automatically propagate changes to the corresponding audio recording without manually editing the original waveform.

A few studies on text-based speech editing make the edited speech more natural.
A pipeline system, called VOCO \cite{jin2018speech},  was completed with the help of a speech synthesis system and voice conversion system. The context-aware prosody correction \cite{morrison2021context} method was further proposed to modify the prosodic information of the target segment to make the prosody of edited speech more natural. To avoid the problems of complicated construction and error accumulation caused by the pipeline system, EditSpeech \cite{tan2021editspeech,tan2022correctspeech} was developed by using partial inference and bidirectional fusion mechanisms. The Alignment-Aware Acoustic-Text Pretraining \cite{bai20223} framework was described to reconstruct masked acoustic signals with text input and acoustic-text alignment during training. In our previous work, a context-aware mask and prediction network (CampNet)  \cite{wang2022campnet} was proposed to
simulate the process of text-based speech editing and can be trained in an end-to-end manner without duration information.

Based on the above series of research on text-based speech editing, the naturalness of edited speech can be guaranteed.
However, some deficiencies will still be in the edited speech, such as dull rhythms and a lack of emotion. Adding an emotional effect to the generated speech is an important measure to eliminate this gap \cite{pierre2003production,erickson2005expressive}. Especially in recent years, with the continuous improvement of generated speech quality, people have increasingly higher requirements for speech style, and emotion-based speech generation is becoming more critical \cite{lorenzo2018investigating, lee2017emotional, wang2018style, kwon2019effective,um2020emotional,zhou2022emotional}. Therefore, in the text-based speech editing, we are going to attempt a new task, which is named "Emotional Selectable Text-based Speech Editing".

 \footnotetext{Examples of generated speech can be found at \href{https://hairuo55.github.io/Emo-CampNet}{https://hairuo55.github.io/Emo-CampNet.}}

To add emotional effects to the editing speech,
we follow the context-aware mask and prediction network \cite{wang2022campnet} to simulate the text-based speech editing process and make the following improvements.   
Firstly, we propose an end-to-end emotion selectable  text-based speech editing network, which can use the input emotion attribute to control the emotion of generated speech. Then, to avoid the emotional components in the generated speech being disturbed by the emotion of the original edited speech, we propose a training framework based on generative adversarial method to extract the emotion-independent content information from the original speech. Finally, we propose two data augmentation methods that can effectively overcome the problems of the small-scale emotional speech dataset and the poor performance of few-shot learning.

Overall, the main contributions of this paper are:
 \begin{itemize}
\item We propose an end-to-end emotion   selectable text-based speech editing model. The key idea of the model is to control the emotion of generated speech by introducing additional emotion attribute based on the context-aware mask prediction framework. 
\item To prevent the emotion of the generated speech from being interfered by the emotional components in the original speech, so that it can only be  controlled by the input emotion attributes,  a neutral content generator is proposed to remove the emotional information from the original speech and is optimized by the generation adversarial training method.
\item Two data augmentation methods  are  proposed to enrich the emotional information and pronunciation information  of the training data, which can effectively improve the model's performance and enable the model to edit any speaker's speech.
\end{itemize}

This paper is structured as follows.  Section \ref{sec:2} introduces the related work. The proposed Emo-CampNet, its operations for emotion selectable text-based speech editing, emotion decoupling with generating adversarial networks, and the data augmentation methods are described in Section \ref{sec:3}. After explaining the experiments and results in Section \ref{sec:4} and Section \ref{sec:5}, we draw a conclusion in Section \ref{sec:6}.
\vspace{-0.1cm}

 

\section{Relate work}\label{sec:2}
This paper is based on the framework of CampNet\cite{wang2022campnet} and adds the emotion selectable function for the generated speech. Therefore, we will briefly introduce the CampNet, then present the work related to emotion speech generation.

\begin{figure}[tp]
    \centering 
    \includegraphics[width=9cm]{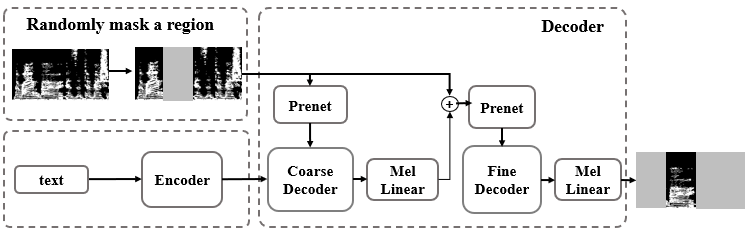}
    \caption{The structure of context-aware mask prediction network (CampNet). To simulate text-based speech editing, the key idea of the model is randomly masking a region of speech and then predicting the masked speech according to the text and speech context.}
    \label{fig:campnet}
\vspace{-0.5cm}
\end{figure}

\subsection{Context-aware mask and prediction network (CampNet)}\label{sec:21}
Generally speaking, there are three operation modes of text-based speech editing: deletion, replacement, and insertion. Although there are many modes of operation, they can be unified into two steps, which are analyzed in detail in \cite{wang2022campnet}. First, mask the region of the original speech that needs to be edited. Then, combine the masked speech and the edited text to predict the masked area. Therefore, the framework of CampNet is designed based on the idea of mask-predict, as shown in Fig. \ref{fig:campnet}. It consists of two processing stages: an encoder and a decoder. First, the encoder module processes the input sentence and converts it into a hidden representation to guide the decoder to predict the acoustic feature of the edited speech. Second, a random region of acoustic features is masked as the ground truth to condition the decoder at the training stage. The decoder is divided into two steps. The first step is to learn the alignment between the masked ground truth and the text representation through the multi-head attention mechanism \cite{vaswani2017attention} and predict coarse acoustic features. Then, the second step is to predict finer acoustic features based on the coarse acoustic features and the original speech context, which can further fuse the context information of speech to make the predicted speech more natural.

\begin{figure}[tp]
    \centering 
    \includegraphics[width=7cm]{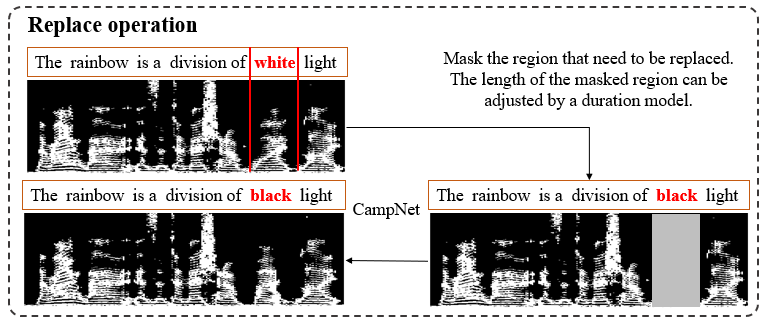}
    \caption{Replacement operation of text-based speech editing  based on CampNet. The operation
can be divided into two steps: first, masking the region to be edited and then
predicting new speech according to the modified text and speech context.
Deletion and insertion operations can also be
divided into similar masking and prediction processes.}
    \label{fig:replace_operation}
\end{figure}

In the test stage, we can use CampNet to complete various operations, such as deletion, operation, and replacement. Here we take the replacement operation as an example, as shown in Fig. \ref{fig:replace_operation}. The first step is to define the word boundary to be replaced, mask it according to the word boundary and then modify the text. The second step is to input the masked speech and the modified text into CampNet. The model will predict the replaced speech according to the modified text. If there is a big difference between the length of the replaced speech and the original speech, such as adding or deleting some words, a pre-trained duration model \cite{wu2016merlin} can be used to predict the length of the replaced region.

\begin{figure*}[htp]
    \centering 
    \includegraphics[width=15cm]{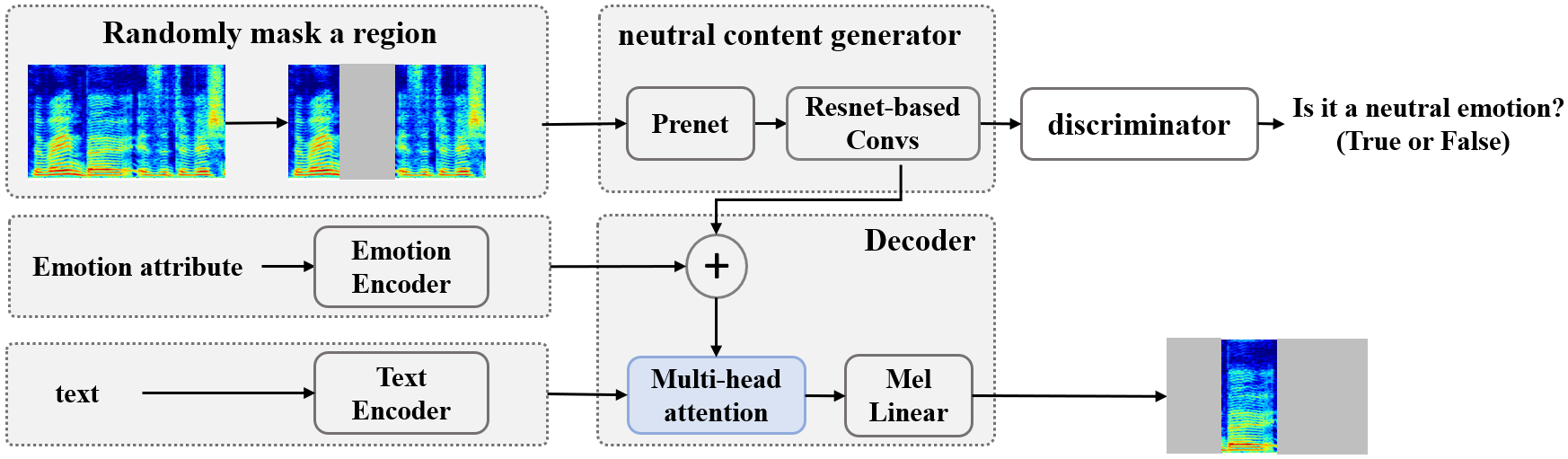}
    \caption{Structure of emotion selectable context aware mask prediction network (Emo-CampNet). Based on the framework of context-aware masking and prediction, emotion attributes are used as input to control the emotion of generated speech. In addition, generative adversarial network is introduced to remove the emotional components from the input speech to prevent the interference to the generated speech.}
    \label{fig:emo-campnet}
\end{figure*}

\subsection{Emotion speech generation}\label{sec:22}

Attempts to add emotional effects to generated speech have existed for more than a decade. Two tasks have been widely studied, emotional text-to-speech \cite{kwon2019effective,choi2019multi} and emotional voice conversion \cite{zhou2022emotional}. For the former task, researchers focused on obtaining expressive emotion representations to guide the system to synthesize expressive speech. Global style tokens (GST) \cite{kwon2019effective} have been proposed to learn the style information in speech, including emotional attributes automatically. In addition, to  more finely control the emotion representation of each phoneme, emotional text-to-speech methods based on the phoneme level are also proposed \cite{lei2021fine,im2022emoq,cui2021emovie}. Some
other studies based on variational autoencoders (VAE) \cite{kingma2013auto}
show the effectiveness of controlling the speech style by
learning, scaling, or combining disentangled representations \cite{zhang2019learning, kenter2019chive}.
For emotional voice conversion, the study focuses on preserving the speech content and converting emotional information. The early studies on emotion voice conversion include GMM and HMM \cite{aihara2012gmm, kawanami2003gmm,zhou2022emotional}. The recent
studies on deep learning have seen a remarkable performance, such as
DNN, sequence-to-sequence models \cite{lorenzo2018investigating, luo2019emotional,luo2016emotional}. With the development of generative adversarial networks (GAN), CycleGAN \cite{9792208, zhou2020transforming,shankar2020non} and StarGAN \cite{rizos2020stargan} have been proposed to disentangle the emotional elements from speech. 

In addition to  two tasks above, this paper attempts a new way to add emotional effects to generated speech: emotion-selectable text-based speech editing, which is presented in the next section.

\section{Emotion-selectable Context-Aware Mask Prediction Network}\label{sec:3}
The difference between the emotion selectable text-based speech editing and the original text-based speech editing is that the former adds a selectable emotion onto the generated speech based on the latter. To implement this function, we first need to introduce additional emotion attribute information to control the emotion of generated speech. Then, we need to decouple the emotion from the edited speech so that the input emotion attributes can better control the generated speech. In addition, to make the model have a better one-shot ability and capable of editing any speaker's speech, we propose two data augmentation methods to enrich the dataset's emotional information and pronunciation information.


In this section, we first introduce the framework of  Emo-CampNet. Secondly, we introduce the  generative adversarial training method for emotion decomposition and control. Thirdly, we introduce the two proposed data agumentation methods.

\subsection{Emo-CampNet and its operations}\label{sec:31}
To make the emotion of the synthesized speech controlled only by the selected emotion attribute and not disturbed by the emotional components in the original speech, we first need to remove the emotion components in the original speech. To remove emotional components from speech, we assume that all emotional speech can be converted to and constructed from neutral emotional speech. Based on the assumption, the emotion selectable speech editing system includes four modules: text encoder, emotion encoder, neutral content generator (NCG), and decoder, as shown in Fig. \ref{fig:emo-campnet}.
Firstly, to realize the basic function of text-based speech editing, we follow the idea of CampNet, i.e., mask part of the speech and then predict the masked area based on the input text and the remaining portions of the speech.
Unlike directly inputting the masked speech into the decoder, a neutral content generator is used to extract the emotion-independent content information to prevent the interference of the emotional information in the original speech to the decoder.
Furthermore, to ensure that the extracted content information is truly independent of emotional information, we use the generative adversarial network to supervise the content information, which will be described in detail in the next section. In the decoding stage, the decoder module predicts the masked area according to the masked neutral content information, text, and emotion attribute. We will describe this process in detail below.

First, given a speech $y$, its text information $ x =  (x_1,\dots, x_{2}, \dots,x_M)$ and  emotional attribute $emo$ --- a one-hot vector corresponding to different emotions (neutral, happy, sad, angry, surprise).  A text encoder processes the input sentence $ x$ and converts it into a hidden representation $h_x$ in the following way:
\begin{equation}
h_x=\left(h_{x_{1}}, h_{x_{2}}, \ldots, h_{x_{M}}\right)=\operatorname{encoder}_{\theta_{x}}(x)\label{eq4}
\end{equation}
where $\theta_{x}$ denotes the parameters of text encoder network.

At the same time, the emotional attribute $emo$ is mapped to a learnable embedding feature, so as to transform the emotional information into the same dimension as the hidden representation of the text $h_x$, this learned embedding can be expressed can be expressed as:
\begin{equation}
h_{emo}=\operatorname{encoder}_{\theta_{emo}}(emo)\label{eq4}
\end{equation}

To simulate the text-based speech editing process, we randomly mask a portion of the speech and then use the network to predict the masked region. In addition, to make the generated speech controlled only by the input emotional attributes $emo$, we need to remove the emotional information from the masked speech. Suppose that the speech after randomly masking a region from $y$ is denoted as $y_{mask}$, the NCG module is used to extract the emotion-independent content information, which can be expressed as:
\begin{equation}
\begin{aligned}
h_{c}=\operatorname{NCG}_{\theta_{c}}(y_{mask})\label{eq5}
\end{aligned}
\end{equation}
where $\theta_{c}$ denotes the parameters of  NCG. The NCG module has two functions. One is to project $y_{mask}$ into the same dimension as the hidden representation of the text $h_x$ and emotion information $h_{emo}$. Second, remove the emotion components  in  $y_{mask}$.


Finally, combined with the hidden features of text $h_x$, emotion $h_{emo}$ and the masked neutral content information $h_{c}$, the masked area of speech is predicted by the decoder, which can be expressed as:
\begin{equation}
y_{pre}=\operatorname{decoder}_{\theta_{d}}(h_x, h_{emo}, h_{c})\label{eq5}
\end{equation}
where $\theta_{d}$ denotes the parameters of decoder network.

\begin{figure}[tp]
    \centering 
    \includegraphics[width=8.5cm]{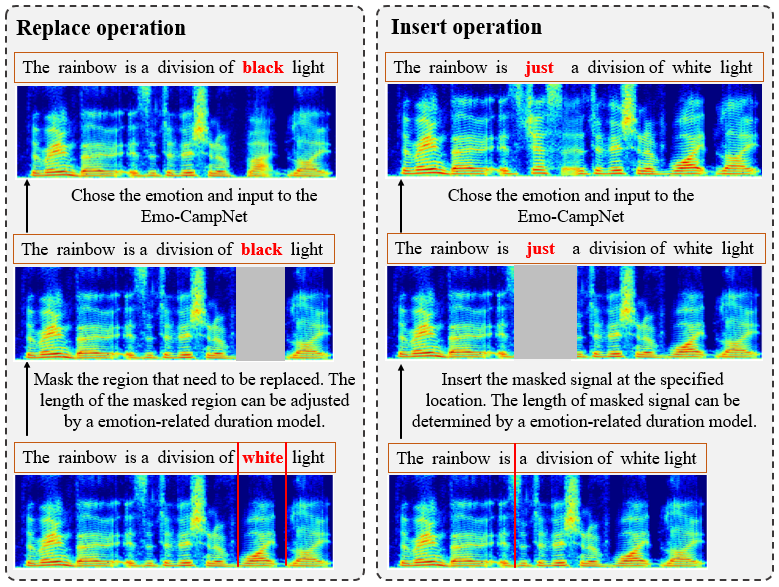}
    \caption{ At the inference stage, the replace and insert operations of emotion selectable text-based speech editing based on Emo-CampNet.}
    \label{fig:emo-campnet-operation}
\vspace{-0.4cm}
\end{figure}

With a pre-trained Emo-CampNet model, some operations of speech editing, such as deletion, insertion, and replacement, can be carried out. These operations are similar to that described in CampNet. Since the deletion operation cannot add any emotional effect, the process is the same as CampNet. Here, we will briefly describe the replacement and insertion operations, as shown in Fig. \ref{fig:emo-campnet-operation}. These operations differ from their counterparts in CampNet is that when we input masked speech and text information into Emo-CampNet, we can additionally select the desired emotional attribute $emo$ of the generated speech, which includes neutral, happy, sad, angry, and surprise. When the Emo-CampNet predicts the speech in the mask area, the speech in this area has the emotional attribute $emo$.

In the decoding stage, ensuring that the extracted content information is independent of emotion is the key to Emo-CampNet. We propose a generative adversarial training method, which will be introduced in the following section.

\subsection{Generative adversarial network for emotion decomposition and control}\label{sec:32}
\begin{figure}[tp]
    \centering 
    \includegraphics[width=7.2cm]{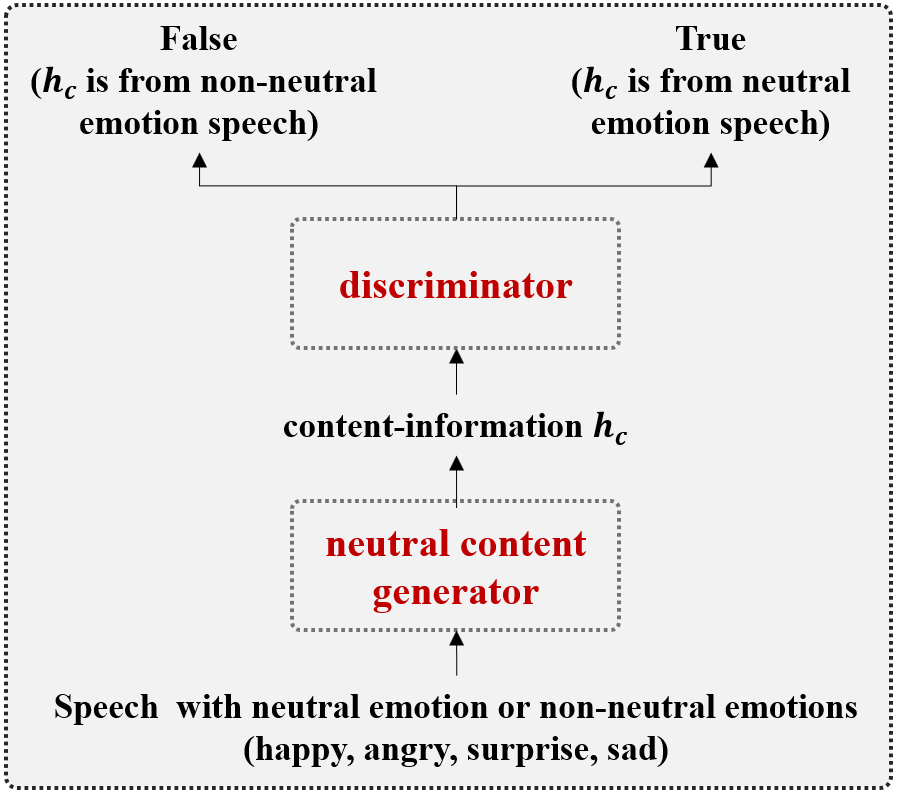}
    \caption{The generative adversarial network is proposed to remove the emotional components in the input speech, which can prevent interference to the emotional information of the generated speech. All the speeches are divided into two categories: neutral emotion speech and non-neutral emotion speech. The discriminator is used to judge whether the content information extracted by the generator comes from neutral speech or non-neutral speech. The generator's task is to confuse the discriminator's judgment.}
    \label{fig:adv}
\vspace{-0.4cm}
\end{figure}

To extract emotion-independent  content information, we assume that all emotional speech can be converted to and constructed from neutral emotional speech. Based on this hypothesis, we propose a generative adversarial training framework \cite{goodfellow2014generative} to remove the emotional components in speech, shown in Fig. \ref{fig:adv}.
Firstly, we use the neutral content generator (NCG) to extract content information $h_c$ from the masked speech $y_{mask}$. To ensure that the content information does not contain emotional components, we introduce a discriminator $D$, where $D$ aims to discern whether $h_c$ is the content information extracted from neutral emotional speech or non-neutral emotional speech. We mark the discriminator target of $h_c$ extracted from neutral emotion speech as $True$ and from non-neutral (happy, angry, surprise, sad) emotion speech as $False$.

Our overall objective contains two components: adversarial losses for transforming emotional speech into neutral content information; and reconstruction loss enables the model to predict the information of the masked area.

\subsubsection{Adversarial loss} \label{sec:adversarial}

We apply adversarial losses to remove the emotional component and retain the content information in speech. For the mapping function $NCG: y \rightarrow h_c $ and its discriminator $D$, we express the objective as:

\begin{equation}
\begin{aligned}
\mathcal{L}_{\mathrm{adv}}\left(NCG, D, Y\right) &=\mathbb{E}_{y \sim p_{\text {neutral }}(y)}\left[\log D(NCG(y))\right] \\
&+\mathbb{E}_{y \sim p_{\text {non-neutral}}(y)}\left[\log \left(1-D(NCG(y))\right]\right.
\end{aligned}
\end{equation}
where $ y \sim p_{\text {neutral }}(y)$ denotes $y$ is neutral emotion speech and $y \sim p_{\text {non-neutral}}(y)$ denotes $y$ is non-neutral emotion speech. $NCG$ tries to generate $h_c$ that is independent of emotion information, while $D$ aims to discern whether $h_c$ contains non-neutral emotional components. When $h_c$ comes from non-neutral emotion speech, the target value of the discriminator is True (denoted with 1). When $h_c$ comes from neutral emotion speech, the target value of the discriminator is False (denoted with 0).  $NCG$ aims to trying to minimize the this objective, while the adversarial discriminator  $D$ tries to maximize it, i.e., $ \min _{NCG} \max _{D} \mathcal{L}_{\mathrm{GAN}}\left(G, D, Y\right) $.

\subsubsection{Reconstruction loss} \label{sec:reconstruction}
By minimizing the adversarial loss, the $NCG$ can generate neutral content information. However, the final goal of the Emo-CampNet is to predict the masked region of acoustic features. Therefore, a reconstruction loss is used in the prediction of the masked region, which is defined as
\begin{equation}
\begin{aligned}
\mathcal{L}_{\mathrm{rec}}\left(y_{pre}, y - y_{mask}\right) = MSE(y_{pre},  y - y_{mask}) \label{recloss}
\end{aligned}
\end{equation}
Where $ y - y_{mask}$ represents the  ground truth of  speech's masked region, and $y_{pre}$ is the speech predicted by the decoder in Eq. \ref{eq5}. $MSE$   stands for mean square error.

\subsubsection{Full objective}
Finally, the objective functions to optimize emo-campnet and discriminator are written, respectively, as

\begin{equation}
\begin{gathered}
\mathcal{L}_{discriminator}=-\mathcal{L}_{adv} \\
\mathcal{L}_{emo-campnet}=\lambda_{adv}\mathcal{L}_{adv}+ \mathcal{L}_{rec}
\end{gathered}
\end{equation}
Where $\lambda_{adv}$ is a hyperparameter that controls the relative importance of the adversarial loss. We use $\lambda_{adv} = 0.5$ in all our experiments.

\subsection{Data augmentation for one shot learning}\label{sec:33}
An important factor limiting the research of emotional speech generation is the small amount of emotional speech data.
Small-scale emotional speech data yields limited emotional, speaker, and pronunciation information, which will make the model prone to overfitting \cite{9229137} and reduce the expressiveness of the model in editing an unseen speaker's speech. Therefore, to improve the model's performance in editing the speech of unseen speakers, it is important to enrich the training data's emotional, speaker, and pronunciation information.
This section proposes two data augmentation methods for the proposed Emo-CampNet model, which can effectively prevent the model from overfitting and improve its expressiveness.


\begin{figure}[tp]
    \centering 
    \includegraphics[width=8.5cm]{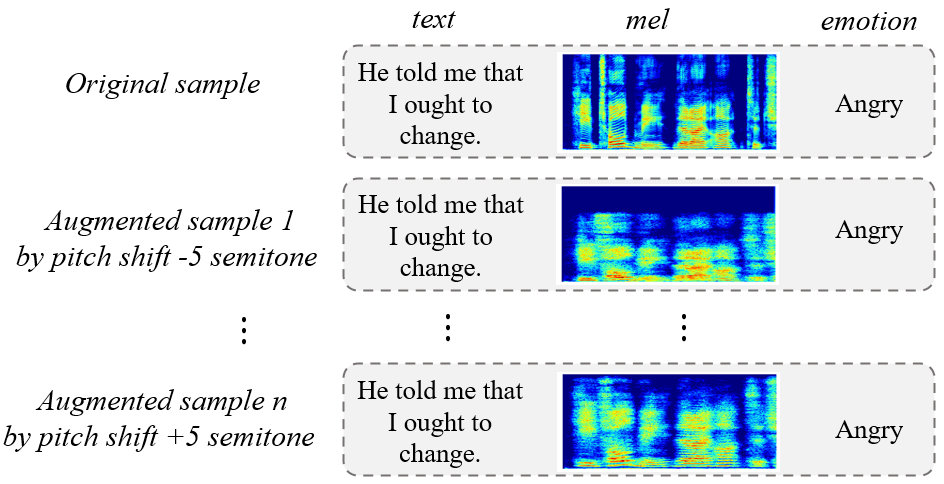}
    \caption{An example of emotion data augmentation. By shifting the F0 of speech, more speech with the same emotion as the original speech can be obtained.}
    \label{fig:data}
\vspace{-0.4cm}
\end{figure}

\subsubsection{emotional data augmentation}
We use the fundamental frequency perturbation method to enlarge the emotional speech data. We disturb the F0 of emotional speech so that the voice color can be changed while its emotional attributes remain unchanged. The data can be augmented $n$ times by $n$ different F0 disturbances on the same utterance. Fig. \ref{fig:data} shows the resulting mel spectrogram excerpts with varying amounts of shifting. The advantage of this method is that it can enrich the data of F0 in different frequency bands without changing the emotional attributes, which helps enrich the training data's emotional and speaker information.


\subsubsection{neutral data augmentation}
Although emotional data can be enriched by F0 disturbance, this method only changes the voice color and does not expand speech content information. To make the model learn more pronunciation information, we use the dataset designed for text-to-speech to expand the training data of Emo-CampNet. This paper uses the open-source dataset VCTK as an example. Since the speech in VCTK is a recording style, we regard its emotional attribute as neutral emotion. In this way, we can introduce many neutral emotional speeches and enrich the speaker and pronunciation information.

This method has two advantages: 1. rich pronunciation information and speaker information can be introduced into the training dataset. 2. Because the number of neutral emotion speech and non-neutral emotion speech is unbalanced when training the discriminator, the data of non-neutral emotion speech is much more than that of neutral emotion speech. This imbalance can be alleviated to some extent by introducing additional neutral emotion data to better train the discriminator.


\section{EXPERIMENTAL PROCEDURES}\label{sec:4}

\subsection{Dataset and Task}
In this section, we conduct experiments on the VCTK  \cite{veaux2017cstr} and ESD \cite{zhou2022emotional} corpora to evaluate our proposed method \footnote{Examples of generated speech can be found at \href{https://hairuo55.github.io/Emo-CampNet}{https://hairuo55.github.io/Emo-CampNet.}}.  The ESD database consists of 350 parallel utterances spoken by 20 speakers and covers 5 emotion categories (neutral, happy, angry, sad, and surprise). The VCTK corpus includes speech data uttered by 110 English speakers with different accents. Each speaker reads out about 400 sentences. Specifically, we select four speakers from the VCTK dataset as the test set, and the remaining utterances are divided into 90\% training set and 10\% validation set. We expand the ESD dataset using the emotion data augmentation method introduced in Section \ref{sec:33}. We take all the training data in VCTK as neutral emotion speech to expand the neutral speech. Finally, we take the training sets of ESD and VCTK as the total training set to train the Emo-CampNet model and take the test set in ESD data and the test set in VCTK as the total test set. All wav files are sampled at 16KHz.



\begin{figure}[tp]
    \centering 
    \includegraphics[width=8.5cm]{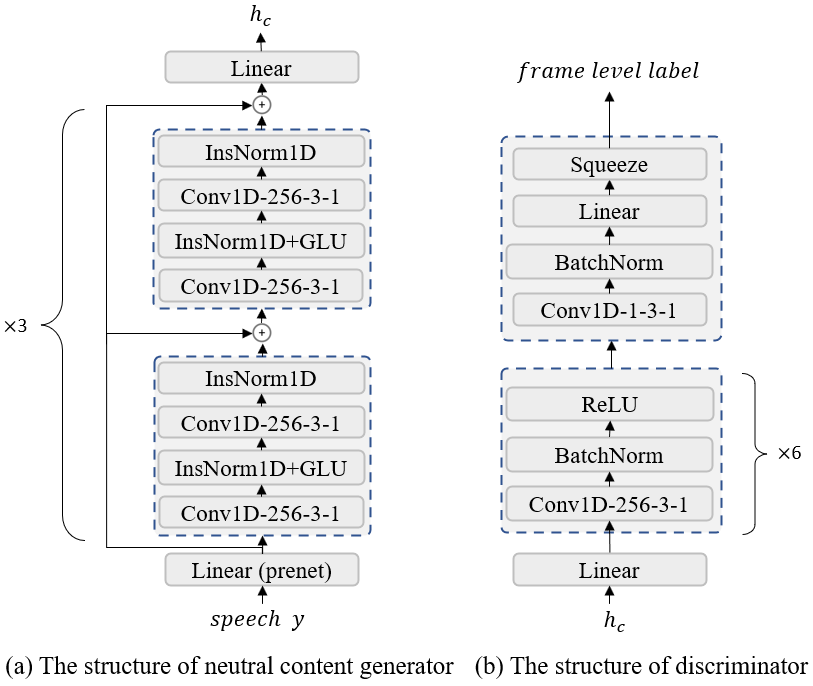}
    \caption{Structures of  neutral content generator and discriminator in Emo-CampNet.}
    \label{fig:structure}
\vspace{-0.4cm}
\end{figure}

\begin{figure*}[tp]
    \centering 
    \includegraphics[width=17.5cm]{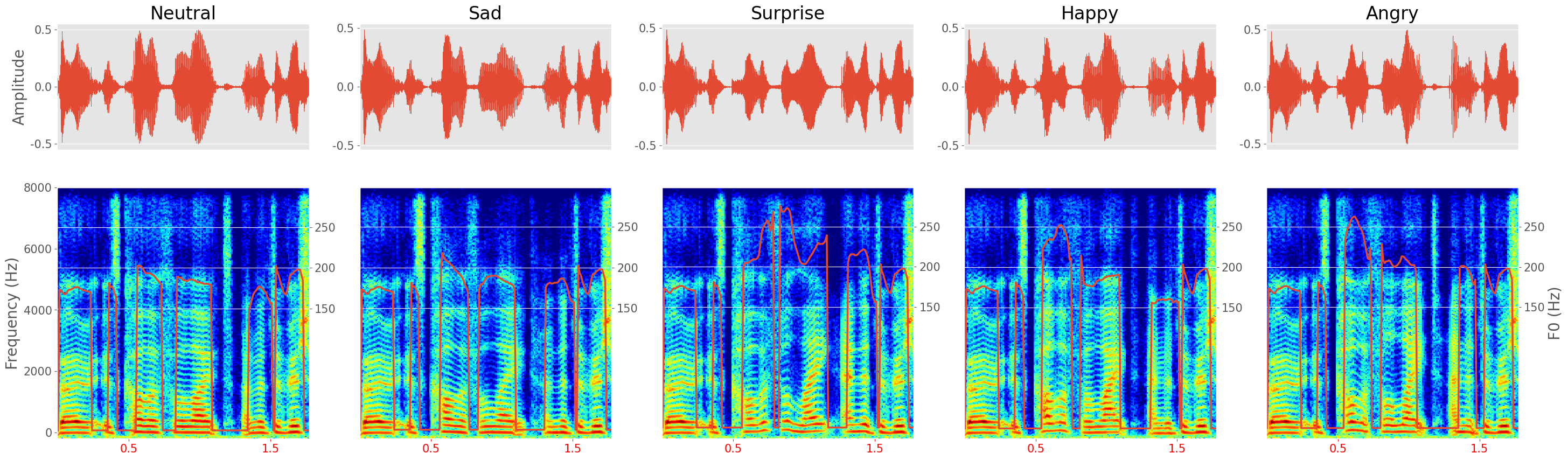}
    \caption{The waveforms and spectrograms of  insertion operation with the same text and different emotion attributes based on the proposed Emo-CampNet. The region marked with time ($0.24s \sim 0.54s$) is the inserted region. The text of the masked region is 'colorful light'.}
    \label{fig:waveform}
\vspace{-0.4cm}
\end{figure*}

\subsection{Model Details}
Acoustic features are extracted with a 10 ms window shift. LPCNet \cite{valin2019lpcnet} is utilized to extract 32-dimensional acoustic features, including 30-dimensional BFCCs \cite{gulzar2014comparative}, 1-dimensional pitch and 1-dimensional pitch correction parameter. Since this is a new task and there has been no relevant work before, we have constructed different baseline systems according to the relevant innovations of this paper. Based on these systems, we can compare the impact of various modules and settings on the performance of the model. There are 5 systems for comparison:
\begin{itemize}
  \item \textbf{Emo-CampNet} First, we train the proposed model Emo-CampNet according to the framework of Fig. \ref{fig:emo-campnet} with the generating adversarial training method and the data augmentation methods. The structure of the neutral content extractor is shown on in Fig. \ref{fig:structure}(a). The structure of the discriminator is shown in Fig. \ref{fig:structure}(b). It should be noted that the output of the discriminator is a frame-level label, which is helpful for the convergence of the model. The decoder and text encoder structures are based on the Transformer structure and are similar to the structure of CampNet. The emotion encoder maps emotion attributes into a learnable embedding. The phoneme sequence is input to a 3-layer CNN to learn the context information of the text. Each phoneme has a trainable embedding of 256 dims. The text encoder and the decoder contain 3 and 6 transformer blocks, respectively.. The hidden dimension of the transformer is 256. We set the masked region to be 12\% of the total speech length at the training stage, which is the same configuration as the CampNet. The initial learning rate is le-3.  Adam \cite{kingma2014adam} is used as the optimizer, with an initial learning rate of $1e-3$. The model is then trained for 2 million steps with a batch size of 16.
  
  \item  \textbf{p-w/o-dis}  The key of Emo-CampNet is removing emotional information from the original masked speech by using the generative adversarial training method. In this way, the generated speech's emotion is controlled only by the input emotion attribute. Therefore, to verify the effect of the generative adversarial method, we remove the discriminator and retrain the model, denoted as p-w/o-dis. The model's loss function is only the reconstruction loss, as shown in Eq. \ref{recloss}. Other configurations are the same as the Emo-CampNet model.

  \item \textbf{p-w/o-ncg} To verify whether the neutral content generator struture we designed is useful for the model, we replace the neutral content generator module with the linear module to extract $h_c$ information. We  keep other experimental configurations unchanged to retrain the model, which is denoted as p-w/o-ncg.
  
  \item \textbf{p-w/o-eda} To verify the impact of the emotional data augmentation method proposed in Sec. \ref{sec:33}, we remove the additional data from the emotional data augmentation method in the training set, keep other experimental settings unchanged, and retrain Emo-CampNet, which is denoted as p-w/o-eda.
  
  \item \textbf{p-w/o-nda}  To verify the impact of the neutral data augmentation method proposed in Sec. \ref{sec:33}, we remove the additional data from the neutral data augmentation method in the training set, keep other experimental settings unchanged, and retrain Emo-CampNet, which is denoted as p-w/o-nda.
\vspace{-0.1cm}
\end{itemize}

\begin{table*}[t]
    \centering
        \caption{Objective evaluation (MCD value) results of different systems on the test set.}
\scalebox{1.2}{
\begin{tabular}{c|ccccc}

\hline \hline   Systems & Neutral & Surprise & Angry &  Happy & Sad \\
\hline   p-w/o-ncg (fail synthesis)  &  - & -  & -  &  - & -  \\
p-w/o-eda  & 3.117  & 3.556   & 3.572  & 3.460    &3.646 \\
p-w/o-nda    &   3.183  & 3.806  & 3.679 & 3.787    & 3.528 \\
  p-w/o-dis    & 3.149 &  3.617     & 3.724  & 3.452   & 3.481 \\ 
Emo-CampNet   & \textbf{3.078} &       \textbf{3.495}     & \textbf{3.528}  & \textbf{3.425}   & \textbf{3.332}  \\

\hline \hline
\end{tabular}}
\label{table:mcd}
\vspace{-0.2cm}
\end{table*}

\section{Results}\label{sec:5}
In this section, we first compare the performance of the proposed Emo-CampNet and some other systems, such as objective metrics and subjective metrics. In addition, since the primary function of Emo-CampNet is the synthesis of emotional speech, we focus on the emotional expressiveness of Emo-CampNet, such as the expressiveness of fundamental frequency and the results of  emotion classification.

\subsection{Objective evaluation for the quality of speech}
First of all, taking the insertion operation as an example, we insert a text at a particular position of a neutral emotional speech, select different emotions as input and synthesize the speech in the corresponding area using the proposed Emo-CampNet. The spectrums of generated speech with different emotions are shown in the Fig. \ref{fig:waveform}. It can be found that although the same text is inserted, the spectrums corresponding to different emotions are very different, and the generated spectrums have natural prosody connection, which shows that the emotion selection function is effective.


Secondly, to objectively compare the performance of different models, we conduct an objective evaluation based on different systems. We take the replacement operation as an example because it is convenient for us to obtain the ground truth of generated speech to calculate the objective metrics.
To prevent the interference of the emotional information in the original speech to the generated speech, we select 20 neutral emotional speeches from the test set in the emotional dataset ESD. Then we randomly select 20 words that span 3 to 10 phonemes from these 20 sentences.
For each sentence, we remove the region of the corresponding words in the speech. Then we use different systems to predict the removed region with the five emotions (neutral, sad, angry, surprise, and happy). Therefore, since each test sentence can be used to synthesize five sentences of the same text but with different emotions, we edited 100 sentences as the test sample to calculate the objective metrics.
Because the ESD dataset contains parallel corpora with different emotions, we can obtain the ground truth of different emotional speech in the removed region.

Following papers \cite{zhou2022emotion, zhou2021limited}, we use the Mel-Cepstral Distortion (MCD) to evaluate emotional speech. The calculation method is as follows:

Given two mel-cepstra $\hat{\mathbf{x}}=\left[\hat{x}_{1}, \ldots, \hat{x}_{M}\right]^{\top}$  and $ \mathbf{x}=\left[x_{1}, \ldots, x_{M}\right]^{\top}$, we use the mel-cepstral distortion (MCD):
\begin{equation}
\mathrm{MCD}[\mathrm{dB}]=\frac{10}{\ln 10} \sqrt{2 \sum_{i=1}^{M}\left(\hat{x}_{i}-x_{i}\right)^{2}}
\end{equation}
to measure their difference. Where  $M$ is the order of mel-cepstrum and  equals 28 in our implementation.   Here, we used the average of the MCDs \cite{kubichek1993mel} taken along the DTW \cite{muller2007dynamic} path between the edited and reference feature sequences as the objective performance measure for each test utterance.

The objective results of the test set are listed in Table \ref{table:mcd}. We elaborate on our conclusions from the following  three aspects.
\subsubsection{Comparison of metric of different emotions}
Firstly, we can find from the results for the Emo-CampNet system that the MCD value of neutral emotional speech is the lowest, and there are similar results in other systems (p-w/o-dis, p-w/o-eda, p-w/o-nda), showing that the acoustic feature of neutral emotion speech is the easiest to learn. This may be because the emotional change of neutral emotional speech is the smallest, and the spectrum is more regular. In addition, the MCD of sad emotion speech is relatively small, which is also because the prosodic change of sad emotion speech is small, and the model is easier to predict. For some emotions with significant prosody changes, such as angry and surprise, their MCD values are greater than other emotions.

 \begin{table*}[tp]
 \centering
 \caption{Average preference score of similarity between generated speech emotion and target emotion attribute of different methods (\%), where n/p stands for "no Preferences", and $p$ represents the $p$ value of the $t$ test
}
 \begin{tabular}{c|ccccc}
 \hline
 \hline
  &  p-w/o-eda  &   p-w/o-dis & Emo-CampNet & N/P & p \\ \hline
p-w/o-eda vs Emo-CampNet   & 9.00  & -- & \textbf{73.00}  & 18.00 & \textless  0.01 \\
p-w/o-dis vs Emo-CampNet     & --   & 6.50  & \textbf{79.00} & 14.50 &  \textless  0.01 \\
 \hline
 \hline
 \end{tabular}
 \label{table:abx}
 \end{table*}

\subsubsection{Comparison with generative adversarial network}
Compared to the p-w/o-dis and p-w/o-ncg model, it can be observed that the two modules (neutral content generator and discriminator) are essential for Emo-CampNet. Specifically, when the Emo-CampNet lacks the NCG module, the model cannot generate natural speech. This is explained by the fact that, without the NCG module, the model cannot extract the speech content information separated from emotion. However, the discriminator always expects the model to generate emotion-independent content information, resulting in the model's collapse.
When Emo-CampNet lacks the discriminator, the model can synthesize neutral-emotion speech with decent fidelity, but its objective measurement results are poor when synthesizing speech with other emotions. This is because with the absence of the discriminator, the emotional information of the speech can not be removed. The emotion of the generated speech is thus affected not only by the input emotion attributes but also by the emotional information of the original speech, resulting in the inability to control the emotion of the generated speech.
\subsubsection{Comparison of data augmentation method}
By comparing model Emo-CampNet with p-w/o-eda and p-w/o-nda models, we can find the influence of data augmentation methods on the objective metric.
When the emotional data augmented by the F0 disturbance is removed from the training data, it can be found that the objective metric is significantly reduced. In particular, the objective metric of non-neutral emotion speech declines more significantly than that of neutral emotion speech.
This is because the fundamental frequency perturbation can change the speaker's information while keeping the emotional attributes unchanged, which is conducive to the expansion of emotional data and the generalization of the model.
Similarly, when the neutral emotion data of VCTK is removed from the training set, the objective metric of p-w/o-nda is significantly reduced because the neutral emotion data set of VCTK can help the model learn richer pronunciation information and speaker information.



\begin{figure}[tp]
    \centering 
    \includegraphics[width=7cm]{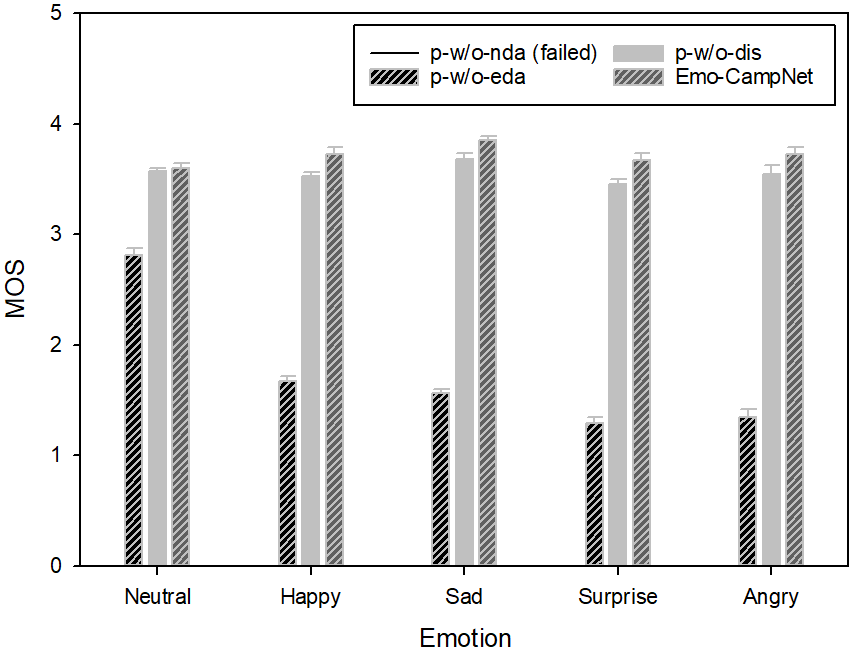}

    \caption{The MOS score with 95\% confidence intervals of the different systems.}
    \label{fig:mos}
\end{figure}

\begin{table*}[t]
    \centering
        \caption{A summary of mean and standard variance (std) of F0 of generated speech with different emotion.}
\scalebox{1.0}{
\begin{tabular}{c|ccccc|ccccc}
\hline \hline    &  &   \multicolumn{2}{c}{female} &  & & & \multicolumn{2}{c}{male}    & &  \\
 \hline Parameter & Neutral   & Surprise  & Angry  & Happy & Sad &  Neutral   & Surprise  & Angry  & Happy & Sad \\
\hline  F0 mean (HZ)  & 197.34  &  242.27  & 211.42  & 219.46  &  200.51 & 130.92  & 176.96  & 155.67   & 169.49 & 136.24  \\
F0 std (HZ)    & 44.50    & 62.42  & 48.13 &  59.19  & 40.37   & 23.06 & 49.14  & 23.43 & 39.62 & 20.73   \\
\hline \hline
\end{tabular}}
\label{table:f0-value}
\end{table*}
 
\subsection{Subjective evaluation for the speech}
In this section, subjective evaluations are conducted to compare the performance of Emo-CampNet and other systems in terms of speech quality and emotional similarity of edited speech. We take the insertion operation as an example. The replacement operation is similar to the insertion operation, except that there are additional steps to mask part of the speech. To prevent the interference of the emotional information of the original speech to the generated speech, we select 40 neutral emotional speeches from the VCTK test set. For each utterance, we insert some words that span 8 to 20 phonemes into these sentences. Then, we use different systems to predict the inserted region with the five emotions (neutral, sad, angry, surprise, and happy). Therefore, each test sentence can be used to synthesize five sentences of the same text but with different emotions. We randomly select 20 parallel sentences for each emotion and collect the mean opinion score (MOS) separately.
Ten listeners are asked to listen and rate the quality and the speaker identity of the edited sentence on a Likert scale: 1 = bad (very bad), 2 = poor (annoying), 3 = fair, 4 = good, and 5 = excellent (imperceptible, almost real). The result is shown in Fig. \ref{fig:mos}.
First, it can be found that p-w/o-nda fails to synthesize speech. This is because much pronunciation information is missing when there is no VCTK dataset in the training set, which combined with the arbitrary text insertion in the test stage, leads to poor performance of the model in the text outside the training set and failure of speech synthesis. In the previous objective evaluation, we have used text data that appears in the training set to obtain the ground truth speech for comparison. The results show that model p-w/o-nda can synthesize speech normally in the text within the training set, but it fails in the text outside the training set.
Secondly, it can be found that model p-w/o-eda performs well in neutral emotion, but the speech synthesized with other emotions is very poor.  The MOS of the Emo-CampNet is slightly higher than that of p-w/o-dis. Since the MOS score mainly evaluates the speech quality and speaker similarity of synthesized speech, we will then evaluate the similarity between the emotion of generated speech and the given emotion attribute.


To this end, we conduct an ABX test to evaluate the emotional similarity of generated speech. In each subjective test, twenty sentences are randomly selected. Ten listeners evaluate each pair of generated speech. The listeners are informed of the target word and emotion attribute in advance, and then asked to pick the generated sample in each pair that is closer to the target emotion, or if they have no preference. The results are listed in Table \ref{table:abx}. Overall, it can be found that Emo-CampNet has a higher score than the baseline system in each evaluation.
In particular, model p-w/o-dis has the lowest score.
After the discriminator is removed, the model cannot predict the speech of a specific emotion according to the input emotion attributes.


\begin{figure}[thp]
    \centering 
    \includegraphics[width=7cm]{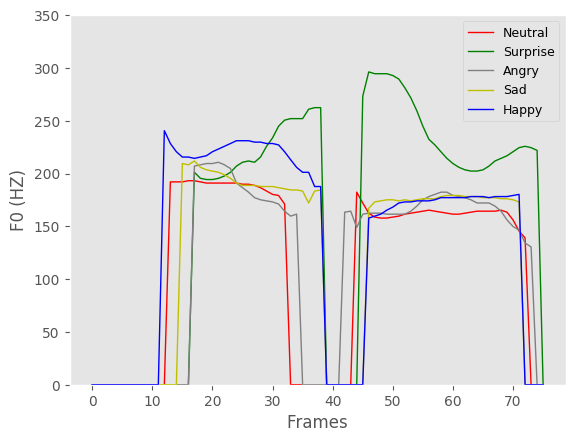}

    \caption{An example of F0 curve  of generated speech with the same text but with different emotion attributes. It can be found that the F0 curve of speech with different emotions has obvious distinction.}
    \label{fig:f0}
\end{figure}

\begin{figure*}[tp]
    \centering 
    \includegraphics[width=16cm]{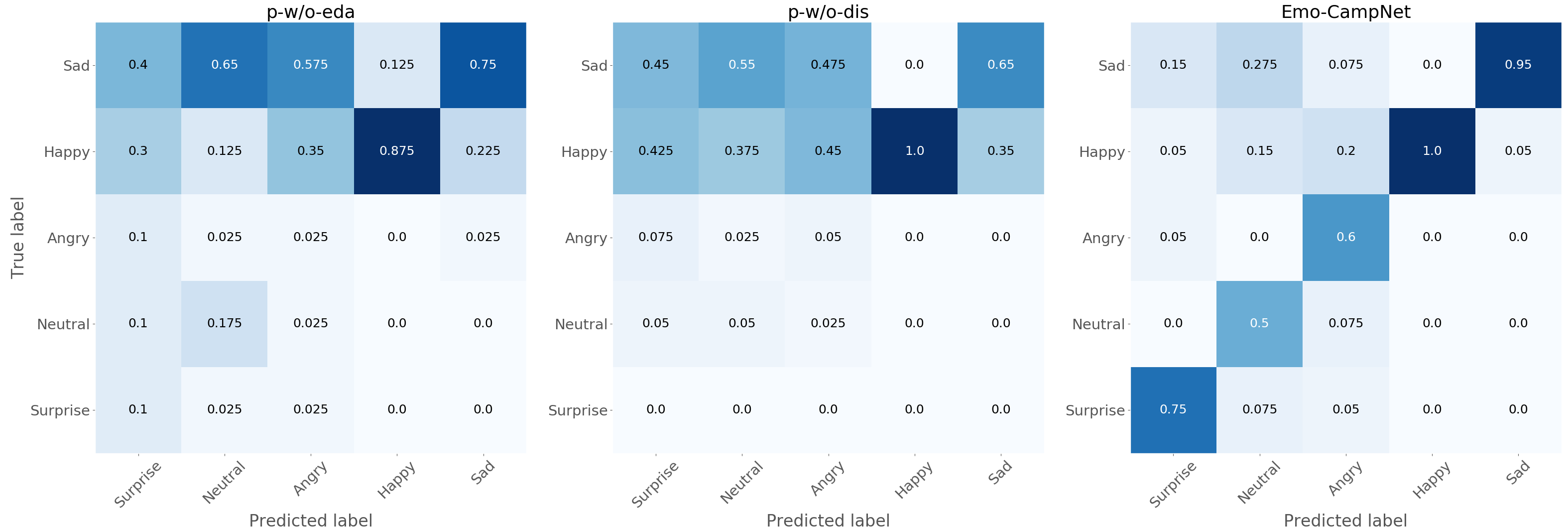}
    \caption{Confusion matrix of different systems in a SER experiment on the test dataset. The diagonal entries represent the recall rates of each emotion.}
    \label{fig:confusion}
\vspace{-0.4cm}
\end{figure*}

\subsection{Expression of fundamental frequency (F0)}

Fundamental frequency (F0) is a critical prosodic descriptor of speech. To observe whether the F0 of speech with different emotions is differentiated, we first display the F0 curve of generated speech by inserting the same text and selecting different emotions.
The F0 value is extracted by the Librosa tool using Yin algorithm \cite{de2002yin}, as shown in Fig. \ref{fig:f0}. It can be found that the F0 of each emotional speech is different. The F0 of neutral emotional speech and sad emotional speech is relatively smooth, while the F0 of surprise, happy and angry emotion speech changes significantly.

Second, we summarize the overall mean and standard variance of F0 in speech editing regions in the test set by emotion and gender. As shown in Table \ref{table:f0-value}, it can be observed that the F0 differs significantly in the mean and variance in synthesized speech with different emotions. The mean of F0 of neutral emotional speech is the lowest, while the F0 of surprise emotional speech has the highest mean value.
In addition, the variance of F0 of sad emotion speech is the smallest, which shows that the change of F0 of sad emotion is the smallest. The variance of F0 of angry emotion speech is the largest, which shows that the F0 of angry emotion speech changes significantly.

\subsection{Emotion classification} 
To validate the quality of emotional expression of generated speech, we follow paper \cite{zhou2022emotional} and develop a speaker-independent speech emotion recognition (SER) model to judge the emotional attributes of synthesized speech. We use the data in ESD as the training set. We use the emotional data augmentation method introduced in Section \ref{sec:33} to augment the training set to improve the SER's performance. We extract the mel spectrum of speech as the acoustic feature, the window size is 25ms, and the hop size of the window is 10ms.

The SER model consists of an LSTM layer, followed by a RELU activated fully connected layer with 256 nodes. Dropout is applied on the LSTM layer with a keep probability of 0.5. Finally, the resulting 256 feature vector is fed to a softmax classifier, an FC layer with 5 nodes. 
We randomly select 40 neutral emotional speeches from the test set. Then we insert some words into these 40 sentences and synthesize them with five emotions. Since the generated speech with long text is more helpful in reflecting the emotional information, the number of words to be inserted is more than 5, and the corresponding speech length is more than 2 seconds. Finally, we input the synthesized speech in the editing area into the pre-trained SER model to judge whether its predicted emotion is consistent with the input emotion attribute.

\begin{table}[t]
    \centering
        \caption{Accuracy of different systems in all emotions in the SER experiment.}
\scalebox{1.3}{
\begin{tabular}{c|c}

\hline \hline   Systems & Accuracy  \\
\hline  
p-w/o-ncg    &   failed synthesis  \\ 
p-w/o-nda    &   failed synthesis  \\
p-w/o-eda  & 0.385  \\
  p-w/o-dis    & 0.35 \\ 
  \hline  
Emo-CampNet   & \textbf{0.76}   \\

\hline \hline
\end{tabular}}
\label{table:confusion}
\vspace{-0.2cm}
\end{table}

The confusion matrices of SER results of different systems are shown in Fig. \ref{fig:confusion}. The value on the diagonal is the proportion that the input emotional attribute is consistent with the predicted emotional attribute. It can be found that model Emo-CampNet has the highest accuracy in each emotion. In addition, it can be found that 27.5\% of the synthesized speech controlled by the neutral emotion attribute is recognized as sad emotion in Emo-CampNet, which indicates that they are easily confused in the synthesis process. This may be due to the small prosodic changes between the two, so there is little difference.

At the same time, it can also be found that when the model loses the discriminator, most of the speech synthesized by the model is recognized as happy and sad emotions. This is because the model is disturbed by the emotional components in the input speech in the training stage after losing the discriminator, so the input emotional attributes can not effectively control the model. In addition, when the two data augmentation methods are removed from the model, the performance decreases significantly, proving our method's effectiveness.

Finally, we compare the accuracy of all emotion recognition of different systems, as shown in Table \ref{table:confusion}. Similar to the results in Fig. \ref{fig:confusion}, it can be found that the Emo-CampNet has the highest accuracy (76\%), which shows that the proposed training framework (generating adversarial networks) and data augmentation methods help improve the performance of the model.




\vspace{-0.1cm}
\section{Conclusion}\label{sec:6}
This paper presents an end-to-end emotion selectable text-based speech editing network, which can control the emotional attributes of synthesized speech in speech editing. This is a new task, and we propose three innovations to realize this function.
Firstly, based on the context-aware mask and prediction framework, emotion attribute information is introduced into the model to guide the decoder to predict speech with a specific emotion.
Secondly, to prevent the emotional components in the original speech from interfering with the decoder, the generative adversarial network is used to remove the emotional components in the original speech. Finally, two data augmentation methods for this task are proposed, which can effectively improve the robustness of the model and the ability of one shot.
The experimental results demonstrate that the proposed method is better than the baseline system in subjective evaluation, objective evaluation, and emotional expressiveness for emotion-selectable text-based speech editing tasks. In addition, ablation experiments also show the effectiveness of our proposed method. Improving the speech quality and expressiveness further is the future work.



%



\ifCLASSOPTIONcompsoc
  \section*{Acknowledgments}
\else
  \section*{Acknowledgment}
\fi
This work is supported by the National Key Research and Development Plan of China (No.2020AAA0140003), the National Natural Science Foundation of China (NSFC) (No.62101553, No.61901473, No.61831022), the Key Research Project (No.2019KD0AD01).

\ifCLASSOPTIONcaptionsoff
  \newpage
\fi



%
\bibliographystyle{IEEEtran}
\bibliography{refs}



%
\begin{IEEEbiography}[{\includegraphics[width=1.1in,height=1.25in,clip,keepaspectratio]{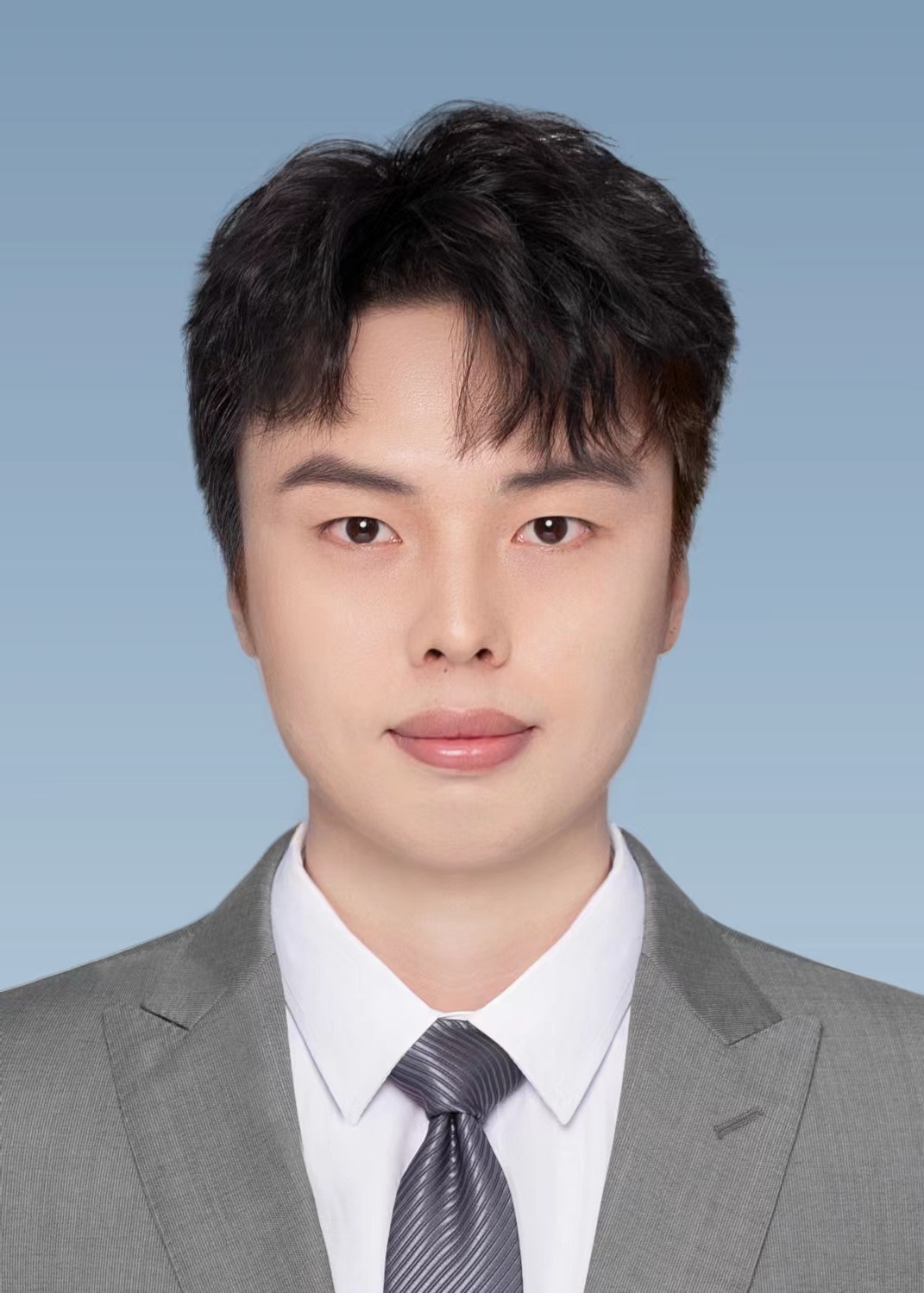}}]{Tao Wang}
received the B.E. degree from the
Department of  Control Science and Engineering, Shandong University (SDU), Jinan, China, in 2018. He is currently working
toward the Ph.D. degree with the National Laboratory
of Pattern Recognition (NLPR), Institute of Automation, Chinese
Academy of Sciences (CASIA), Beijing, China. His current
research interests include speech synthesis, voice conversion,  machine learning, and transfer learning.
\end{IEEEbiography}
\vspace{-1.0cm}

\begin{IEEEbiography}[{\includegraphics[width=1.1in,height=1.25in,clip,keepaspectratio]{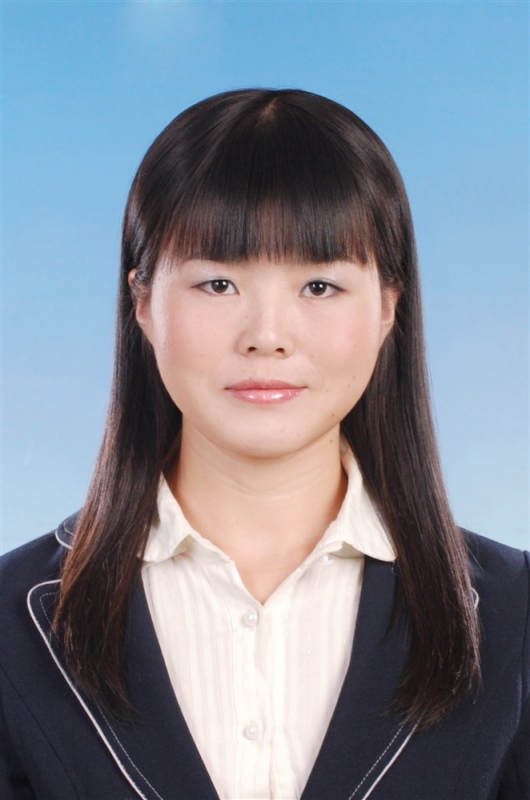}}]{Jiangyan Yi}
received the Ph.D. degree from the
University of Chinese Academy of Sciences, Beijing,
China, in 2018, and the M.A. degree from the Graduate
School of Chinese Academy of Social Sciences,
Beijing, China, in 2010. She was a Senior R\&D
Engineer with Alibaba Group from 2011 to 2014.
She is currently an Assistant Professor with the National
Laboratory of Pattern Recognition, Institute of
Automation, Chinese Academy of Sciences, Beijing,
China. Her current research interests include speech
processing, speech recognition, distributed computing,
deep learning, and transfer learning.
\end{IEEEbiography}
\vspace{-1.0cm}

\begin{IEEEbiography}[{\includegraphics[width=1.1in,height=1.25in,clip,keepaspectratio]{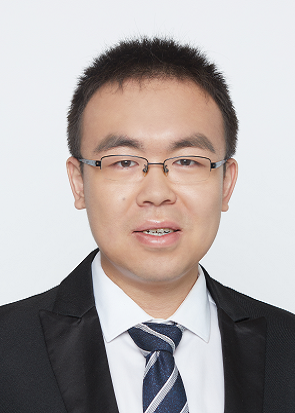}}]{Ruibo Fu}
is an assistant professor in National
Laboratory of Pattern Recognition, Institute of Automation, Chinese Academy
of Sciences, Beijing. He obtained B.E. from Beijing University of Aeronautics and Astronautics in 2015 and Ph.D. from Institute of Automation, Chinese Academy of Sciences in 2020. His research interest is speech synthesis and transfer learning. He has published more than 10 papers in international conferences and journals such as ICASSP and INTERSPEECH and has won the best paper award twice in NCMMSC 2017 and 2019.
\end{IEEEbiography}
\vspace{-1.0cm}

\begin{IEEEbiography}[{\includegraphics[width=1.1in,height=1.25in,clip,keepaspectratio]{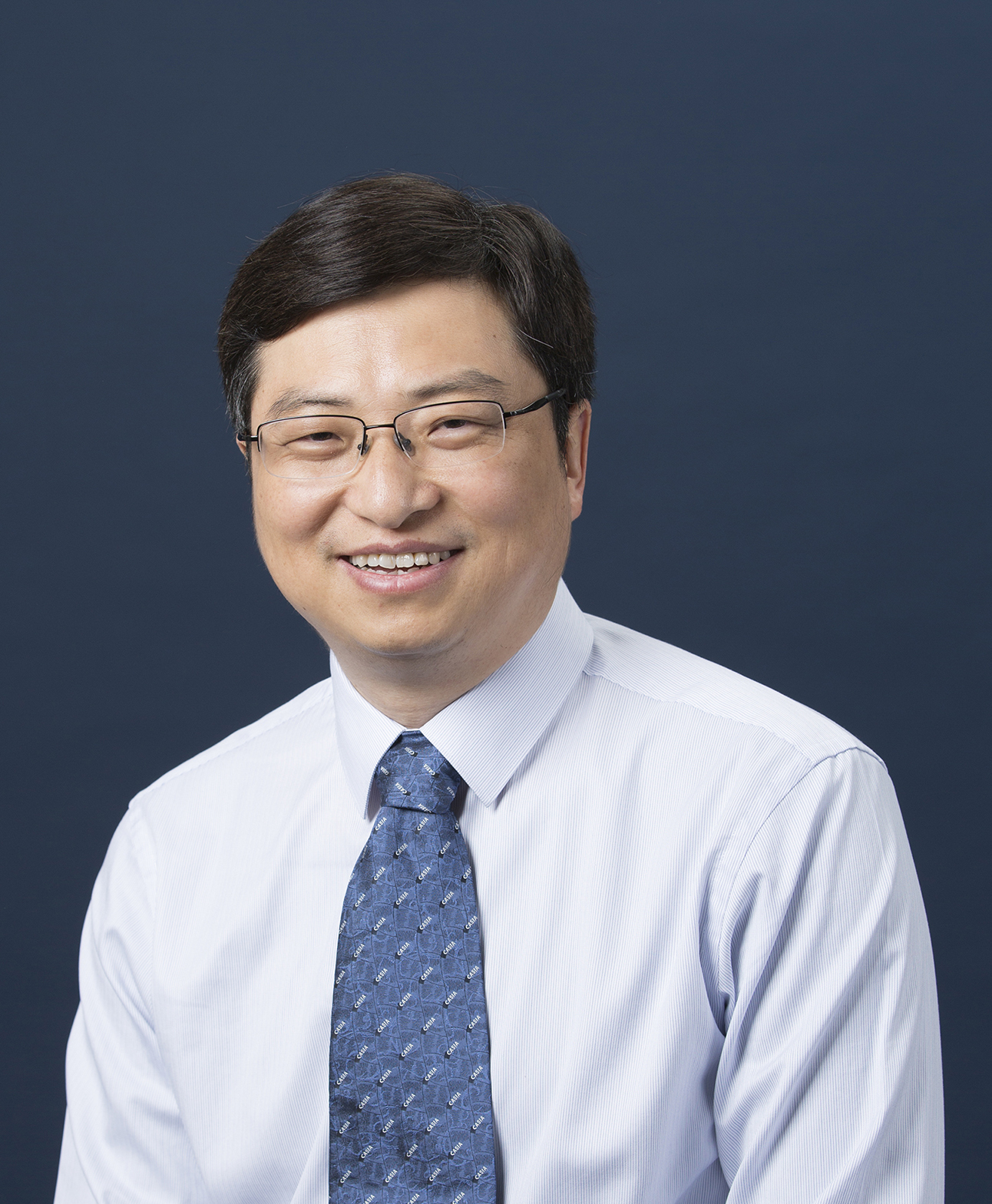}}]{Jianhua Tao}
(SM’10) received the Ph.D. degree from
Tsinghua University, Beijing, China, in 2001, and
the M.S. degree from Nanjing University, Nanjing,
China, in 1996. He is currently a Professor with
NLPR, Institute of Automation, Chinese Academy
of Sciences, Beijing, China. His current research interests
include speech synthesis and coding methods,
human computer interaction, multimedia information
processing, and pattern recognition. He has authored
or coauthored more than 80 papers on major journals
and proceedings including IEEE TRANSACTIONS ON
AUDIO, SPEECH, AND LANGUAGE PROCESSING, and received several awards
from the important conferences, such as Eurospeech, NCMMSC, etc. He serves
as the chair or program committee member for several major conferences,
including ICPR, ACII, ICMI, ISCSLP, NCMMSC, etc. He also serves as the
steering committee member for IEEE Transactions on Affective Computing, an
Associate Editor for Journal on Multimodal User Interface and International
Journal on Synthetic Emotions, the Deputy Editor-in-Chief for Chinese Journal
of Phonetics.
\end{IEEEbiography}
\vspace{-1.0cm}

\begin{IEEEbiography}[{\includegraphics[width=1.1in,height=1.25in,clip,keepaspectratio]{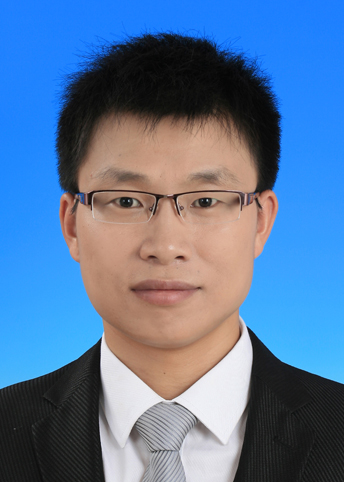}}]{Zhengqi Wen}
received the B.E. degree from the
Department of Automation, University of Science and
Technology of China, Hefei, China, in 2008 and the
Ph.D. degree from the National Laboratory of Pattern
Recognition, Institute of Automation, Chinese
Academy of Sciences, Beijing, China, in 2013. From
March 2009 to June 2009, he was an intern student
with Nokia Research Center, China. From December
2011 to March 2012, he was an intern student
with the Faculty of Systems Engineering,Wakayama
University, Japan. From July 2014 to January 2015,
he was a visiting scholar, under the supervision of Professor Chin-Hui Lee,
with the School of Electrical and Computer Engineering, Georgia Institute of
Technology, USA. He is currently an Associate Professor with the National
Laboratory of Pattern Recognition, Institute of Automation, Chinese Academy
of Sciences, Beijing, China.
\end{IEEEbiography}

\begin{IEEEbiography}[{\includegraphics[width=1.0in,height=1.25in,clip,keepaspectratio]{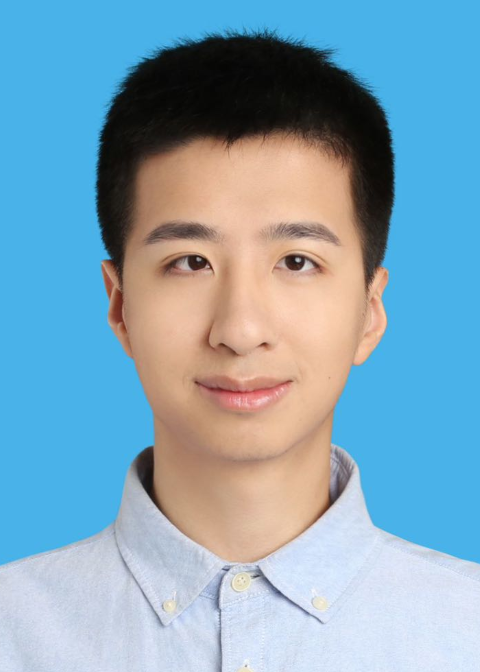}}]{Chu Yuan Zhang}
 received his B.A. degree in linguistics at the University of California, Los Angeles (UCLA), in 2021. He is currently pursuing a M.E. degree in Pattern Recognition and Intelligent Systems with the International College,  National Laboratory of Pattern Recognition (NLPR) at the Institute of Automation, Chinese Academy of Sciences (CASIA). His current research interests include speech synthesis and machine learning.
\end{IEEEbiography}



\end{document}